\documentclass{jps-cp}
\usepackage{txfonts} 

\usepackage{siunitx}
\usepackage{url}

\title{Efficient Analysis of Test-beam Data with the\\ Corryvreckan Framework}

\author{Jens \textsc{Kr\"oger}$^{1,2}$ and Lennart \textsc{Huth}${^3}$ \\
	on behalf of the CLICdp collaboration and the Corryvreckan developers}

\inst{
$^{1}$Physikalisches Institut, Universit\"at Heidelberg, Im Neuenheimer Feld 226, 69120 Heidelberg, Germany \\
$^{2}$CERN, Espl.~des Particules 1, 1211 Meyrin, Switzerland \\
$^{3}$DESY, Notkestraße 85, 22607 Hamburg, Germany
}

\email{kroeger@physi.uni-heidelberg.de}

\recdate{November 18, 2020}

\abst{Stringent requirements are posed on the next generations of vertex and tracking detectors for high-energy physics experiments to reach the foreseen physics goals. A large variety of silicon pixel sensors targeting the specific needs of each use case are developed and tested both in laboratory and test-beam measurement campaigns. Corryvreckan is a flexible, fast and lightweight test-beam data reconstruction and analysis framework based on a modular concept of the reconstruction chain. It is designed to fulfil the requirements for offline event building in complex data-taking environments combining detectors with different readout schemes. Its modular architecture separates the framework core from the implementation of reconstruction, analysis and detector specific algorithms. In this paper, a brief overview of the software framework and the reconstruction and analysis chain is provided. This is complemented by an example analysis of a data set using the offline event building capabilities of the framework and an improved event building scheme allowing for a more efficient usage of test-beam data exploiting the pivot pixel information of the Mimosa26 sensors.}

\kword{software architectures (event data models, frameworks and databases); analysis and statistical methods; data processing methods; pattern recognition, cluster finding, calibration and fitting methods; solid state detectors}

\begin{document}
\maketitle

\section{Introduction}
\label{sec:intro}
The physics objectives of future high-energy particle physics experiments pose stringent requirements on the tracking detector systems.
With the advances of the semiconductor industry, silicon pixel sensors are an increasingly competitive technology capable to meet these requirements.
Not only the planned upgrades of the experiments at the Large Hadron Collider like ATLAS~\cite{atlas-itk} or CMS~\cite{cms-phase2}, but also other possible future high-energy physics experiments, for example at the Compact Linear Collider (CLIC)~\cite{yellow-report-summary}, foresee the use of silicon pixel sensors in their vertex and tracking detector systems.
As a consequence, R\&D on a vast range of both hybrid and monolithic pixel detector technologies is carried out.

New detector prototypes are often pushing the limits of technology well beyond what has proven to work in the past.
Hence, it is crucial that simulations and laboratory measurements are complemented by test-beam campaigns, in which pixel detectors are tested and characterised with a beam of minimum ionising particles.
Beam telescopes provide a precise reference measurement of particle tracks, which can be used to determine performance parameters of the pixel detectors under investigation such as the hit detection efficiency as well as the spatial and timing resolution.

Due to the large variety of applications, different readout architectures are implemented depending on the particular requirements of each use case.
This creates new challenges regarding the integration of new sensors into test-beam data acquisition systems, as well as for the reconstruction and analysis software, which needs to be capable of combining data from detectors with different readout schemes and data structures in order to avoid repeated implementations of similar algorithms.

The \textit{Corryvreckan} framework~\cite{corry-website,corry-git} addresses these challenges and provides a flexible and highly configurable tool for the efficient analysis of test-beam (and laboratory) data.
It was initially developed within the CLIC Detector \& Physics (CLICdp) collaboration~\cite{clicdp-website, yellowreport-detector} and is now employed and extended by a number of users and developers from different experiments supporting an increasing variety of sensor prototypes.
A number of publications underline the versatility and the success of the framework~\cite{pitters-tpx3,bugiel-soi,dort-proceedings,kroeger-proceedings,munker-proceedings,williams-proceedings,phd-quast,phd-williams}.
Through its modularity, it aims to maximize synergies between different research groups by providing a reusable and extendable code basis and thus suppressing the need for numerous similar frameworks.
Its offline event building scheme allows for a combination of data from detectors with different readout schemes in multiple ways depending on the analysis objectives~\cite{corry-paper}.
It is developed with the goal to be easy to use and understand, and provides comprehensive documentation~\cite{corry-manual-v1.0}.

This paper gives an overview of the framework architecture and briefly describes the reconstruction and analysis flow (Sect.~\ref{sec:framework}).
In Sect.~\ref{sec:example}, an example analysis is presented showing the capabilities of the offline event building and illustrating a recent improvement for a more efficient data analysis.
Finally, Sect.~\ref{sec:conclusion} provides a conclusion and an outlook to future developments.

\section{The Corryvreckan Framework}
\label{sec:framework}
\textit{Corryvreckan} is designed following a \textbf{modular architecture} comprising a framework core and user modules.
The core handles all centrally provided functionality.
This includes the user interaction via the command line interface as well as the parsing of configuration files, data input and output, and internal coordinate transformations.
It is separated from modules dedicated to specific reconstruction tasks, such as reading of raw data, clustering, or track reconstruction.
Owing to this modularity, users can add their own functionality to the reconstruction chain by developing new modules targeting their particular needs.

The software is configured with the help of two text-based files.
The main configuration file defines the reconstruction and analysis chain by listing the required modules and providing the relevant parameters.
In addition, a geometry configuration file is needed, in which the position and rotation of all detectors in the setup and their properties are defined.

\begin{figure}[tbh]
	\centering
	\includegraphics[width=1.0\textwidth]{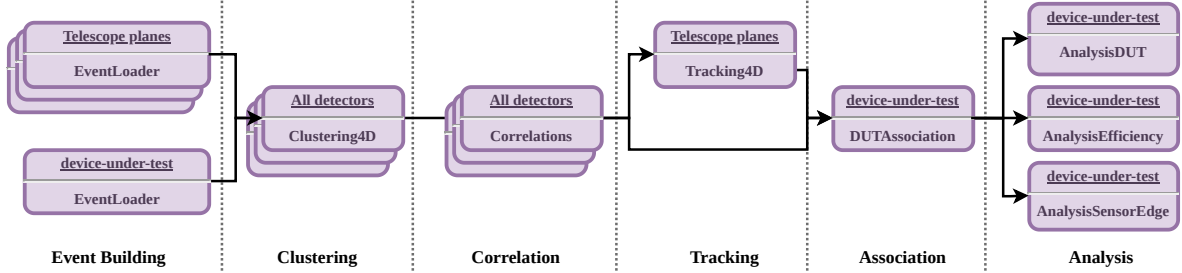}
	\caption{Flow chart of a reconstruction and analysis chain focussing on the performance of the DUT. Multiple modules for different analysis objectives are included. Modified from~\cite{corry-paper}.}
	\label{fig:reco-chain-telescopeWithDUT}
\end{figure}

As mentioned above, the flexible \textbf{reconstruction and analysis chain} in \textit{Corryvreckan} can be adjusted according to the objectives of the analysis as will be shown in Sect.~\ref{sec:example}.
The steps performed during the reconstruction are defined by the modules listed in the main configuration file.
Figure~\ref{fig:reco-chain-telescopeWithDUT} shows a full reconstruction and analysis chain as it can be used to investigate the performance of a device-under-test (DUT).
The individuals steps are briefly outlined below.

\paragraph{Event Building}
When decoding raw data from different detectors and converting them into pixel hits, these need to be arranged in coherent data chunks containing the relevant data from all detectors in order to facilitate a consistent analysis.
This is referred to as \textbf{event building} in \textit{Corryvreckan}.
An event is defined by a time interval with a start and an end time.
In addition, it may contain unique trigger identifiers (trigger IDs) and trigger timestamps.

The first module of the reconstruction chain needs to define the extent of an event.
Depending on the type of detector, an event definition can be derived from data, e.g.~for a shutter-based device.
Otherwise, a module called \texttt{Metronome} can be used to create events of user-defined length.
Each subsequent detector can only add matching data to this event, either based on timestamps or by comparing trigger IDs.
This task is performed by modules called \texttt{EventLoaders}.
A more detailed description of different event building schemes can be found in~\cite{corry-paper}.
One example is discussed in Sect.~\ref{sec:example} of this paper.\\

\paragraph{Clustering}
Depending on the incident point and angle of a traversing ionising particle, electron-hole pairs can be created in two or more adjacent pixels.
In addition, charge can be detected in neighbouring pixels e.g.~due to lateral diffusion.
This effect is called charge sharing.

Hence, pixel hits originating from the same traversing ionising particle need to be grouped into so-called clusters.
Depending on the available information, a clustering algorithm based on spatial information (\texttt{ClusteringSpatial}) or spatial and time information (\texttt{Clustering4D}) can be applied.\\

\paragraph{Correlations}
Spatial and time correlations can be calculated between clusters on each detector plane and a user-defined reference plane.
These are particularly helpful to check data quality and ensure synchronisation and alignment between devices.
In addition, they can be used for a coarse pre-alignment prior to a precise track-based alignment (see below).
They are provided by the \texttt{Correlations} module.\\

\paragraph{Track finding and track fitting}
Clusters on the telescope planes need to be combined to form reference tracks.
For the track finding, candidate clusters are identified on each telescope plane.
In a second step, a track model such as a straight-line track or a General Broken Lines (GBL)~\cite{gbl-fitting} model is fitted through these clusters.
This task can be performed by different \texttt{Tracking} modules in \textit{Corryvreckan}.\\

\paragraph{Association of DUT clusters with reference tracks}
For an unbiased analysis, the device-under-test should be excluded from the tracking procedure.
Consequently, clusters on the DUT need to be associated with the reference tracks based on spatial and time cuts in a separate step of the reconstruction done by the \texttt{DUTAssociation} module.\\

\paragraph{Track-based alignment}
In addition to a coarse pre-alignment of the setup based on the correlations as described above, \textit{Corryvreckan} supports a track-based alignment of the reference telescope by minimising the global track $\chi^2$ using Minuit2~\cite{minuit2} or the Millepede-II~\cite{millepede} algorithm.
For the device-under-test, which should be excluded from the tracking, an alignment is possible via the minimisation of its unbiased track-hit residuals. \\

\paragraph{Analysis of performance parameters}
In a final step, different analysis modules can be used to investigate particular performance parameters of the reference telescope and the DUT, such as the hit detection efficiency or the spatial and time resolution.

\section{Combining Data from Different Devices by Offline Event Building}
\label{sec:example}
Data from different detectors can be combined using the offline event building capabilities of \textit{Corryvreckan}.
In this section, this is illustated by an example based on data recorded at the DESY II Test Beam Facility~\cite{desy-ii}.

\subsection{Experimental Setup}
\begin{figure}[tbh]
	\centering
	\includegraphics[width=0.8\textwidth]{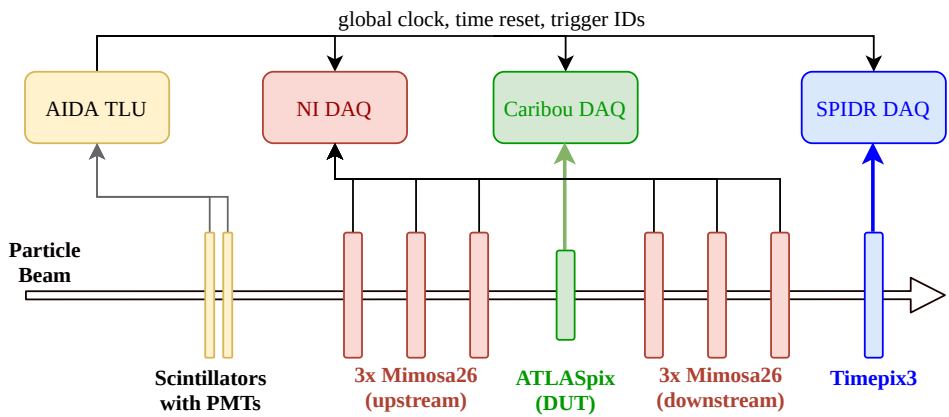}
	\caption{Schematic drawing of the experimental setup.}
	\label{fig:setup}
\end{figure}

A schematic drawing of the experimental setup is shown in Fig.~\ref{fig:setup}.
The data acquisition (DAQ) was controlled via the EUDAQ2 software framework~\cite{eudaq2} using the following devices:

\begin{itemize}
	\item The AIDA \textbf{Trigger Logic Unit (TLU)}~\cite{aida-tlu} (shown in yellow) distributes a global clock and a global time reset, and receives input signals from two scintillators. 
	When a coincidence is detected, it generates a trigger ID, records trigger timestamp and ID, and sends out the trigger ID to all connected devices.
	\item The EUDET-type \textbf{reference telescope}~\cite{eudet-telescopes} consists of six monolithic Mimosa26 sensors (shown in red).
	They are read out by the NI DAQ system in a continuous rolling shutter mode with a periodicity of \SI{115.2}{\micro s}.
	Only when a trigger signal is received, the data of the corresponding shutter period are stored. 
	The trigger is provided by the TLU and the associated trigger ID is recorded together with the data to allow for an offline synchronization with the other detectors.
	Individual pixel timestamps are not available.
	Hits from Mimosa26 planes and the Timepix3 (see below) are combined into reference tracks.
	\item As an additional \textbf{timing reference plane} a Timepix3 hybrid silicon pixel detector assembly~\cite{timepix3} (shown in blue) is operated.
	Its precise pixel timestamps are used to assign unambiguous track timestamps.
	As a data-driven detector it is always active and sends its pixel hits off to the SPIDR DAQ system~\cite{spidr} directly after detection without the need of an external trigger signal.
	\item The \textbf{device-under-test (DUT)} of the presented setup is an ATLASpix~\cite{atlaspix-peric} high-voltage monolithic active pixel sensor (shown in green). 
	Like the Timepix3 it is a data-driven detector, which is always active and sends its hit data including pixel timestamps off to the Caribou DAQ system~\cite{caribou} immediately after detection.
\end{itemize}

Since the telescope, the timing reference plane, and the DUT are read out using three independent DAQ systems, the data streams do not have an inherent global event definition.
This fact will be important in the following analysis example. The capability of combining these data streams in the offline analysis is one of the core strengths of the \textit{Corryvreckan} event building logic.

\begin{figure}[bh]
	\centering
	\includegraphics[width=0.9\textwidth]{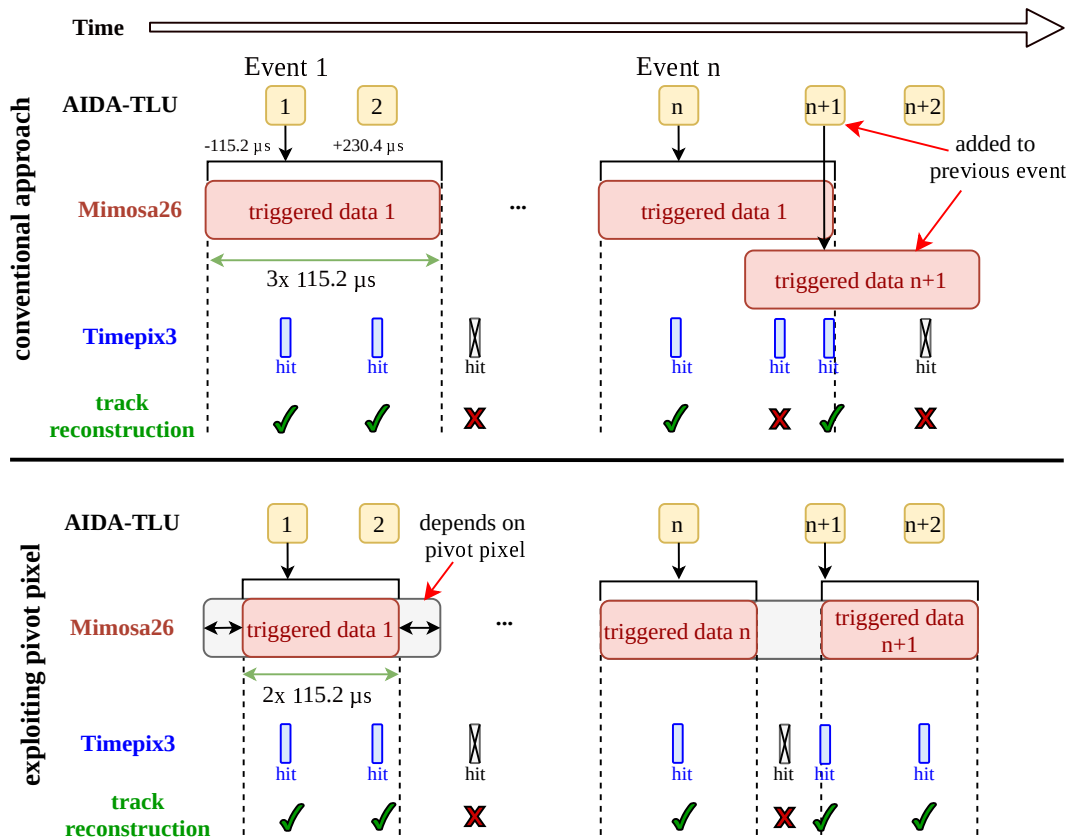}
	\caption{Schematic of the conventional event building (top) and the improved event building (bottom) exploiting the pivot pixel information of the Mimosa26 detectors.}
	\label{fig:eventbuilding-tlu-m26-tpx3}
\end{figure}

\subsection{Conventional event building with the AIDA-TLU and the Mimosa26 sensors}
\label{subsec:analysis-telescope}
As the experimental setup consists of six triggered devices and two always-active data-driven detectors, a reasonable event definition should be based on the active time of the triggered devices.
Outside of these active time windows, no hits are recorded on the Mimosa26 planes so that no reference tracks can be reconstructed.
Since the Mimosa26 data do not contain timestamps but only trigger IDs, their active time must be reconstructed using the TLU relating the trigger IDs to timestamps.
Hence, the event building needs to start with the TLU defining an event with a trigger timestamp and a trigger ID, so that the Mimosa26 data can be matched by comparing the trigger ID.
The event can now be refined by spanning a time interval around the trigger timestamp covering the time in which the Mimosa26 hits may have been recorded.
According to the design of the rolling shutter readout scheme with a periodicity of \SI{115.2}{\micro s}, this time span can cover a maximum of one cycle (\SI{115.2}{\micro s}) before and two cycles (\SI{230.4}{\micro s}) after the trigger time.
The exact interval depends on the position of the rolling shutter at the moment when a trigger signal is received.
In the conventional approach, which has been widely used, e.g.~in~\cite{kroeger-proceedings}, a fixed time window corresponding to the largest possible time interval is used as illustrated at the top left of Fig.~\ref{fig:eventbuilding-tlu-m26-tpx3}.
Thereafter, hits from the data-driven detectors Timepix3 and ATLASpix are added to the event based on individual pixel timestamps, i.e.~hits with timestamps earlier than the currently defined event are discarded, those with a later timestamp are kept for the next event, ensuring that a coherent chunk of data is used for the analysis.

Occasionally, the over-estimation of the active time of the device can lead to an additional trigger being added to the current event.
If this additional trigger actually happened after the readout of the current event and the Mimosa26 sensors are active again, then this data block is also added to the current event by matching the trigger ID as illustrated on the top right of Fig.~\ref{fig:eventbuilding-tlu-m26-tpx3}.
Since the Timepix3 data are added based on its pixel timestamps and a hit on this detector is required during the tracking, no tracks can be found for the additional hits on the Mimosa26 planes from the additional readout block.
In the next event, the Mimosa26 data block has been used, such that potential tracks from these hits cannot be reconstructed.
This effect causes a slight reduction in the number of tracks that are reconstructed during analysis for a given data set.
However, because the track reconstruction is restricted to the well-defined time span of the event by requiring a Timepix3 hit as part of the track, the analysis of the DUT performance is not compromised.

\begin{figure}[b]
	\begin{minipage}[t]{.45\linewidth}
		\centering
		\includegraphics[width=\textwidth]{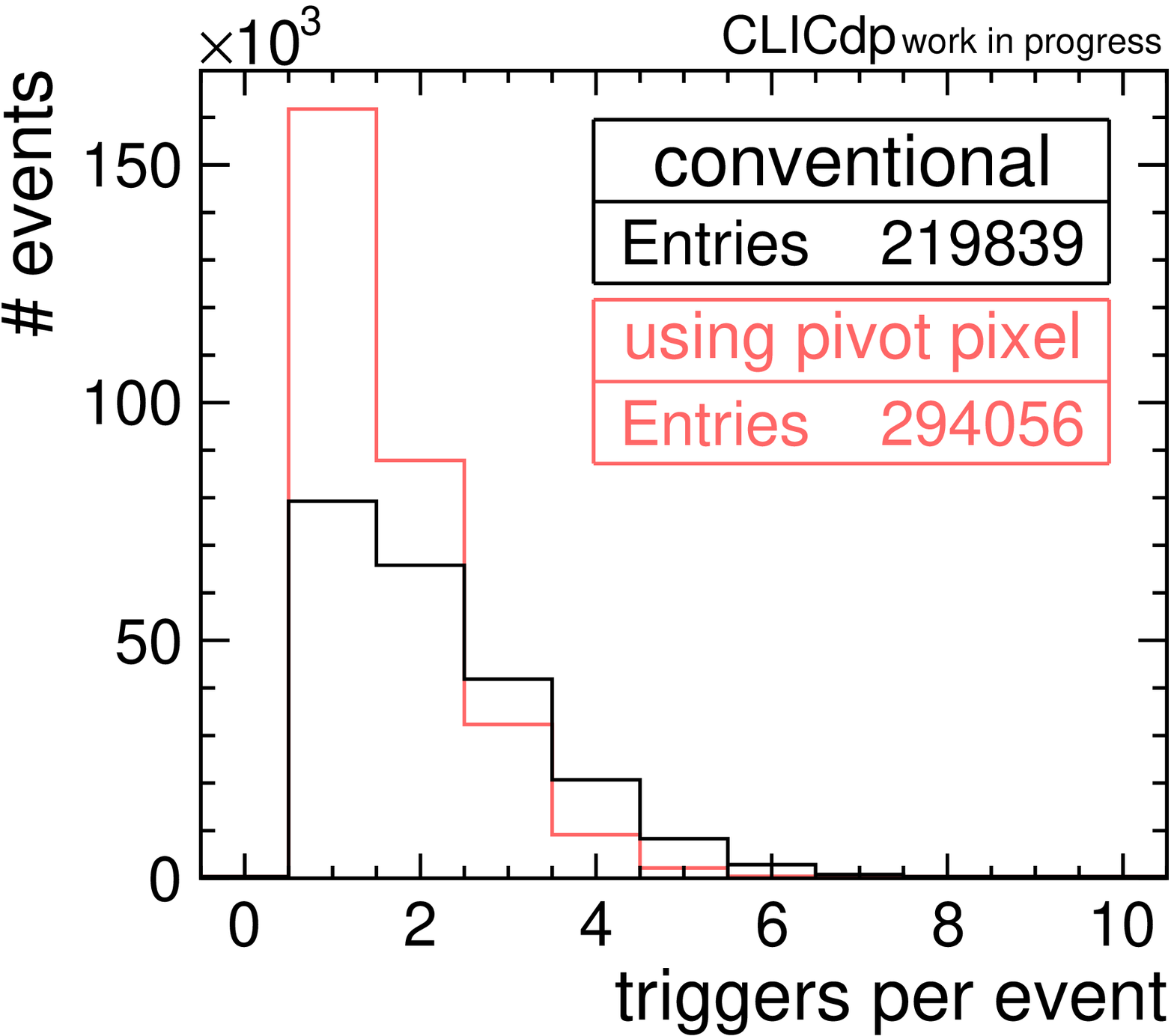}
		\caption{Number of triggers per event for both the conventional and the improved event building exploiting the pivot pixel for the same data set with \SI{600}{s} of data taking.}
		\label{fig:triggersPerEvent}
	\end{minipage}%
	\hfill
	\begin{minipage}[t]{.45\textwidth}
		\centering
		\includegraphics[width=\textwidth]{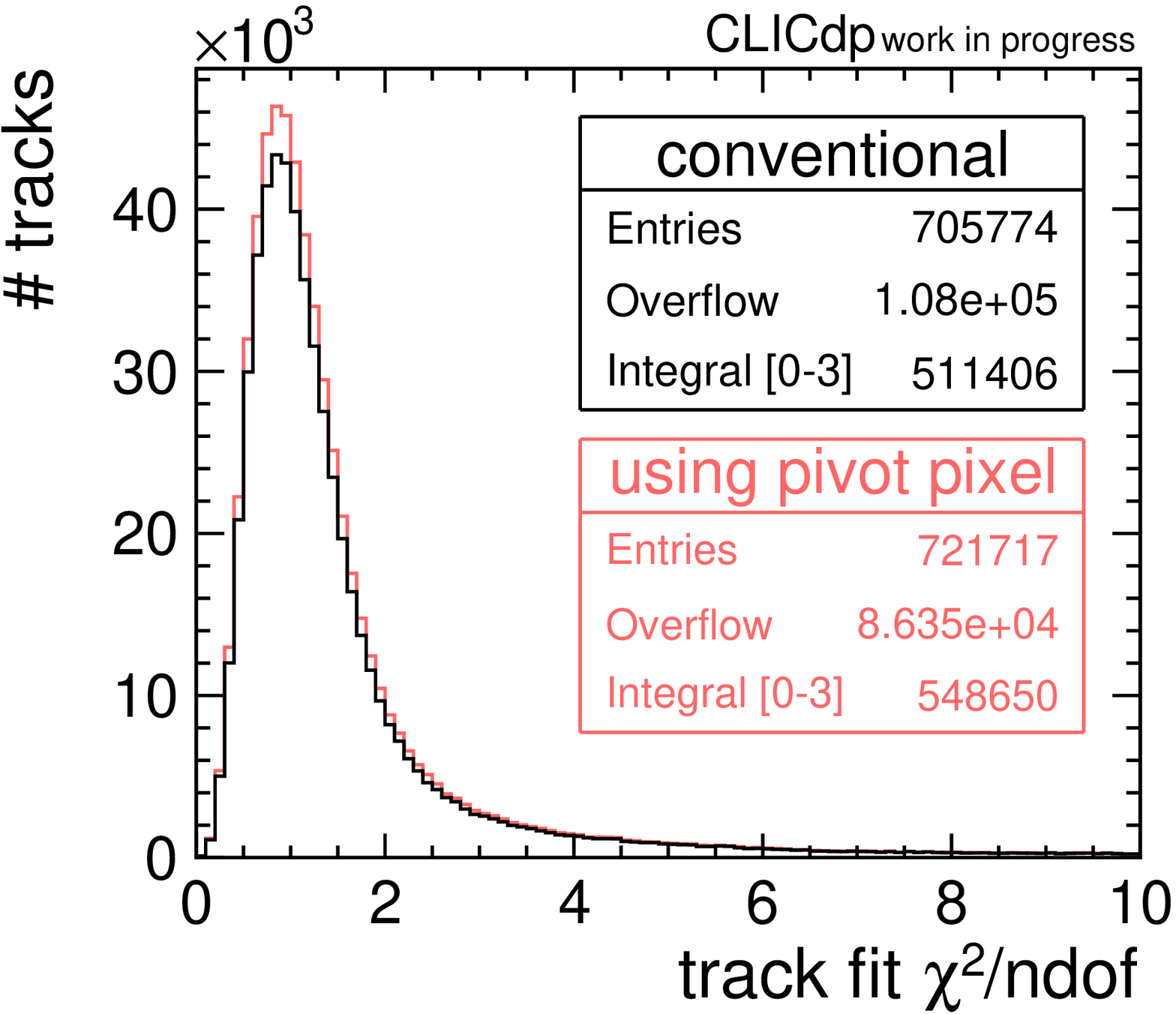}
		\caption{Track fit $\chi^2$ divided by the number of degrees of freedom ($n_{\mathrm{dof}}$) for both the conventional and the improved event building exploiting the pivot pixel for the same data set with \SI{600}{s} of data taking.}
		\label{fig:trackChi2}
	\end{minipage}
\end{figure}

\subsection{Improved event building exploiting the Mimosa26 pivot pixel information}
In order to improve the usage of the data, a new module \texttt{EventDefinitionM26}~\cite{corry-git-evdefm26} has been developed.
It makes use of the fact that the Mimosa26 data contains the information which pixel row the rolling shutter was passing when a trigger signal was received, the so-called \textbf{pivot pixel}.
Using this information, the precise time interval with a duration of two readout cycles (\SI{230.4}{\micro s}) can be calculated relative to the trigger timestamp as indicated at the bottom left of Fig.~\ref{fig:eventbuilding-tlu-m26-tpx3}.
Consequently, the event length is reduced from \SI{345.6}{\micro s} to \SI{230.4}{\micro s}.
As can be seen in Fig.~\ref{fig:triggersPerEvent}, this increases the number of reconstructed events in the investigated run by \SI{\approx 34}{\percent} as expected from the ratio of event lengths, whereas the number of triggers per event is reduced.
If the number of triggers per event is now larger than one, it means that the TLU has recorded an additional coincidence and this trigger is stored. 
But because the Mimosa26 readout is still busy at this point in time, no corresponding block of data can be found for these trigger IDs and thus no second block of Mimosa26 data is added to the event.

Figure~\ref{fig:trackChi2} shows the distribution of the track fit $\chi^2/n_{\mathrm{dof}}$ (number of degrees of freedom), which is a measure for the quality of the fit.
Typically, tracks with a $\chi^2/n_{\mathrm{dof}} < 3$ are used for the final analysis.
In the analysed data set with a data taking time of \SI{600}{s}, the number of tracks with a $\chi^2/n_{\mathrm{dof}} < 3$ is increased by \SI{\approx 7}{\percent} compared to the conventional approach meaning that the number of tracks usable for the analysis is enhanced.
This is a consequence of the more precise event time definition such that Timepix3 hits, which have previously been discarded, are now added to the additional events as illustrated in the bottom right of Fig.~\ref{fig:eventbuilding-tlu-m26-tpx3}.
However, the number of reconstructed tracks does not increase by \SI{34}{\percent} like the number of events because also the conventional approach is not restricted to one track per event. 
As illustrated in the top left of Fig.~\ref{fig:eventbuilding-tlu-m26-tpx3}, more than one track can be found and reconstructed in one event even if the second trigger is received during the readout block associated to trigger ID 1.
This is due to the architecture of the Mimosa26 readout: If a second trigger occurs during the readout initiated by the first trigger, additional particles can still detected if they hit the pixel matrix in front of the rolling shutter. They are then ("wrongly") associated to trigger ID 1 even though they were caused by particle 2.
But since all hits including the Timepix3 hit are added to the same event, the track can be reconstructed.

Simultaneously, the number of tracks with a large $\chi^2/n_{\mathrm{dof}} > 50$ (shown in the overflow bin) is reduced significantly by \SI{\approx 20}{\percent} due to the fact that no additional Mimosa26 data blocks are added to an event and thus the combinatorics of track candidates is reduced significantly.
These tracks do not play a role for the further analysis because they are rejected based on their large $\chi^2/n_{\mathrm{dof}}$.
However, they slow down the computation time needed for the analysis because of the larger combinatorics while calculating correlations and performing the track finding and fitting.

By applying this improvement in the event building logic, existing test-beam data can be exploited more efficiently because fewer data are discarded and more usable tracks are reconstructed for a given data set.
The improvement becomes more significant when taking data with a high particle rate, where multiple triggers per event occur even more often.
This is, for instance, the case for lower beam momenta at DESY~\cite{desy-ii}:
At \SI{2}{GeV/c} the beam rate is about ten times higher than at \SI{5.4}{\GeV/c}, which was used for the data presented in this publication.
It is also particularly valuable for small devices-under-test, on which the number of particle tracks is limited due to their geometrical dimensions such that a full exploitation of the recorded data is essential to get the highest possible statistics in the analysis.
Most importantly, the usage of the \texttt{EventDefinitionM26} module is crucial for a correct determination of the hit detection efficiency of a device-under-test in a EUDET-type reference telescope in which an additional timing reference plane, such as the Timepix3, is not available.
Only with the presented improvement, the correct active time of the reference telescope can be reconstructed.

\section{Conclusion \& Outlook}
\label{sec:conclusion}
To pursue the foreseen physics goals of future experiments in high-energy physics, a broad spectrum of silicon pixel detectors is being developed.
The large variety of technologies requires a new level of flexibility of the reconstruction and analysis software tools for test-beam data allowing to combine data from devices with different readout schemes.

The \textit{Corryvreckan} framework addressed these needs by providing a flexible and highly configurable test-beam data reconstruction and analysis toolkit.
It has a modular architecture separating the framework core from modules for specific reconstruction tasks.
An example has been presented showing how a data set can be analysed using the flexible event building feature of the framework.
In addition, a recent improvement in the event building logic exploiting the pivot pixel information for the Mimosa26 sensors has been shown.
This leads to an increase in the number of usable reconstructed reference tracks that can be reconstructed from a given data set and allows for a correct hit detection efficiency determination even if no additional timing reference plane is available.

Due to its modularity and the high level of documentation, \textit{Corryvreckan} enables collaborative development and long-term maintenance of its code basis shared between different research groups.
Driven by the specific needs in various test-beam analysis projects, the software is extended by its users and new modules integrating further detectors, tracking in magnetic fields, and an extension allowing multi-threading are currently under development.

\section*{Acknowledgements}
This work has been sponsored by the Wolfgang Gentner Programme of the German Federal Ministry of Education and Research (grant no. 05E15CHA).
The measurements leading to these results have been performed at the Test Beam Facility at DESY Hamburg (Germany), a member of the Helmholtz Association (HGF).

\end{document}